# An equation for the isosbestic point wavelength in aqueous solutions of electrolytes

De Ninno A.[1*], De Francesco M.[2]

[1] ENEA UTAPRAD-DIM, C.R. Frascati, Via Enrico Fermi, 45 00044 Frascati (Roma) Italy

[2] ENEA PROMAS-MATPRO C.R. Casaccia, Via Anguillarese 301, S. Maria di Galeria 00123 (Roma) Italy

[*] antonella.deninno@enea.it; +39 0694005792

# ABSTRACT

The existence of an isosbestic point in the OH bond vibration band of $H_2O$ molecules in solutions of strong electrolytes has been shown by many research groups but its physical meaning has been only recently read as the equilibrium point between two populations of water molecules where either water-ion or water-water interactions respectively prevail. Thus the isosbestic point provides useful information on the bulk displacement of water molecules in presence of solutes. Herein we show that the isosbestic point occurrence is accurately described by an equation depending on the balance of the electrostatic forces of the ion and its counter-ion. Furthermore, its wavelength is also related to the slope of the density vs. solute concentration. This support the view that the effect of the solute can be barely regulated within few hydration shells.

## 1. INTRODUCTION

It has been thoroughly discussed among scientists whether solutes may also have a long-range restructuring effect on water [1] or if they only modify the first sphere (or spheres) of hydration [2]. Classical theories of the electrolytes predict that water molecules form an electric shield around each ion dissolved into the solution. The amount of molecules sized by the solute are actually under debate because of the difficulties in univocally defining the ability of the ions to tight bond the water. Depending on the definition given to the adjective "tight", different kinds of hydration number can be obtained by different experimental techniques [3]. Hydration numbers obtained by the colligative properties are usually lower than those obtained by diffraction techniques (X-rays, neutron scattering). The latter techniques measure the number of molecules surrounding the ion, thus depending on the ion size, and the results coincide with the Molecular Dynamics (MD) simulations. However, the colligative properties, which are relevant to the solute-solvent interaction strength, depend on the ions charge density. As an example, in case of the ion $Cl^-$, the cryogenic lowering gives a dynamic hydration number ≈ 0 [4] whilst scattering measurements found as many as six water molecules surrounding one $Cl^-$ ion [5].

A useful model to discuss the effect of solvation from the point of view of the solvent is the concept of Debye-Hückel cages, which are defined as the regions occupied by the solvent that is heavily affected by the presence of the solute (cations and anions). At low concentrations of the solute, water cages do not overlap, however, when two solute particles come close together enough, the molecules inside can, in principle, either relax to the state of the normal bulk solvent or give rise to a mutual constructive interaction. In a previous paper our observations have shown that the long-range structure of water, on increasing the concentration of the solute, arranges essentially in three different phases: a low concentration phase, where the solutes are accommodated without important distortion of the water matrix; an intermediate phase, where a strong cooperative behaviour is observed among water molecules, and a third, where the deformation reaches its maximum value because any further increase of the ions in solution is prevented by the recombination. It can be shown that, by drawing the ratio of the areas (under the absorption spectra) below and above the isosbestic point, their ratio describes a sigmoid function. In particular we are dealing with a Hill function whose rate is independent on the nature of the solute and it is just linked to the bulk properties of the solvent [6]. Such an observation shed a new light on the hydration process of strong electrolytes and provides, for the first time, a theoretical basis for the description of high concentration electrolytic solutions out of the Debye-Hückel validity range.

We have also shown that the isosbestic point $\omega_{i.p.}$ marks the difference between two populations of water molecules characterised by different intensity of the water-water and water-ions interactions. It can be observed by the FTIR spectra that the solutes induce a deformation in the OH stretching band of water. In particular, for $\omega < \omega_{i.p.}$ the band drifts towards high energy, whilst for $\omega > \omega_{i.p.}$ we can observe a red shift. This is due to a deformation of the OH bond length suggesting a different distortion of the dipole moments of water molecules. In case the electrostatic attraction between water molecule and ion is prevailing the dipole is stretched, on the contrary whenever the long-range water-water mutual interaction prevail the water dipole is shortened [7].

The isosbestic, equal-absorption point emerges as the equilibrium point between two fractions of water molecules whose structure is differently affected by the solute. It represents the equal absorption frequency $\omega_{ip}$ of the two fractions that doesn't change with the concentration of the solute [8, 9]. Water molecules that have a stretching vibration lower than $\omega_{ip}$ are related to high-correlation, low density molecules, whilst on the other hand, low correlated molecules (small cluster, dimers) are responsible for O-H stretching at frequency higher than $\omega_{ip}$ [10, 11]. We thus asked: can the occurrence of the isosbestic point be exactly predicted from the features of the ions ?

## 2. EXPERIMENTAL AND METHODS

The vibrational spectra of the solutions have been acquired by the FTIR spectrometer Nicolet iS5 equipped with a diamond crystal ATR iD7 from ThermoScientific. Chemical products used to prepare the solutions have a purity > 99% and have been used as received to prepare aqueous 1m solutions. Bi-distilled water has been produced with a quartz bi-distiller and housed at 20° in Teflon flasks.

All the measurements were made at 20°C with thermostatic control of the sample. The spectra have been obtained as the average on 16 acquisitions in the range 400-4000 $cm^{-1}$, the background subtraction procedure have been done before each acquisition. Each spectrum has been compared with the water spectrum obtained in the same experimental conditions.

## 3. RESULTS AND DISCUSSION

In the following we will describe how the study of the isosbestic points of the electrolytic solutions actually provides a large amount of information about the characteristics of the solutions and, in particular, about the water molecules organization. We have looked mainly at the following issues:

a. the occurrence of isosbestic points on the scale of the wavelength follows the order: $Cs^+ > Rb^+ > K^+ > Na^+ > Li^+$ for each halide, which coincides with the Hofmeister's series [12]. The key parameter of this series is the charge density of the cations and of the corresponding anions. Another key parameter, for the appearance of the isosbestic point, is the electronegativity of the anions. These two parameters, combined in an equation, let us to predict with a good accuracy the wavelength of the isosbestic point provided the knowledge of both anion and cation of mono and bi-valent alkali halides or, vice versa, to detect the ionic radius of an unknown cation by knowing its isosbestic point. The equation can also be extended to other cations such as, for example, $NH_4^+$.

b. so far the density change of an electrolytic solution has been only associated to the substitution of a solvent with a solute molecules. Being the isosbestic point related to the ratio between the low-density and high-density fraction of water, we should expect a relationship between the isosbestic point of the solution and the density. We also report a formula allowing to calculate the slope of the density increase with the solute with knowing the isosbestic point of the solution.

c. halogenic acids also show an isosbestic point, however it is worthwhile to ask whether it has the same physical meaning as in case of salts. We suggest that, even if the isosbestic point has a different physical origin, it can be used in evaluating the effective ionic radius of the $H_3O^+$ in aqueous solution.

## AN EXACT EQUATION FOR THE ISOSBESTIC POINT

The OH stretching broad band of aqueous solutions, in the range 2800-3800 $cm^{-1}$, is strongly affected both by the type and by the concentration of the solutes. In particular, it is known that an isosbestic (equal absorption) point appears by increasing the molar concentration. This means that the overall spectrum can be decomposed exactly into the sum of the two components, one absorbing at $\omega < \omega_{i.p.}$ and another with stretching vibrations $\omega > \omega_{i.p.}$. In spectroscopic notation, the abscissa normally goes from higher (left) to lower (right) wavenumbers, thus low-energy vibrations appear on the i.p. right-hand side and are related to highly-correlated low-density molecules whereas the left-hand side is related to poorly correlated having higher density. The isosbestic point sharply divides water molecules with respect to the effect of solutes on their OH (intramolecular) bond length.

Isosbestic points have been measured for 1 m alkali halides at 20°C, the results are shown in Table I and, by way of example, the occurrence of isosbestic points for bromides is shown in Fig. 1. FTIR spectra of the remaining three halides and the bi-valent chlorides are shown in the supplemental material.

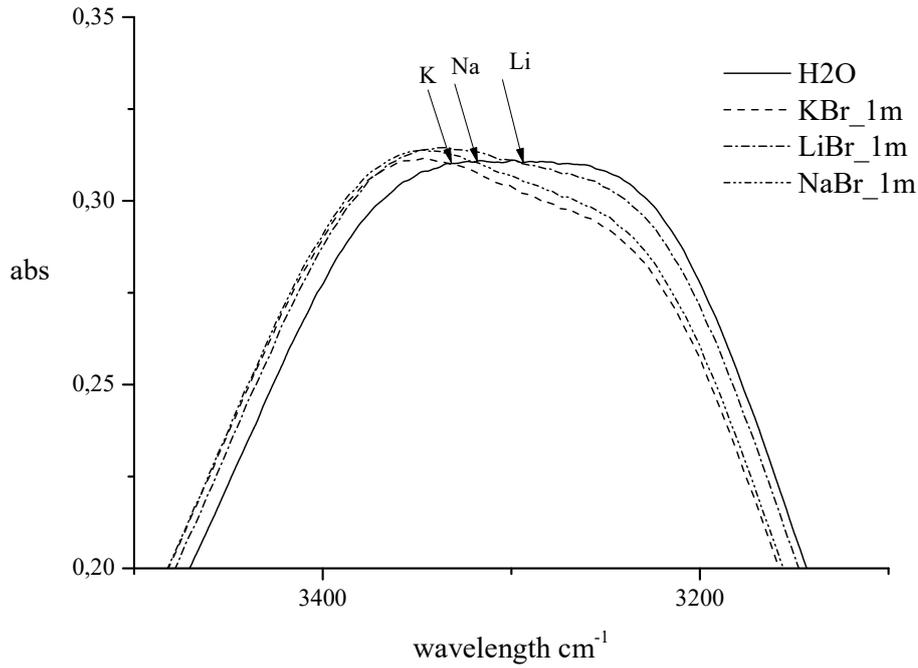

Fig.1 FTIR spectra of 1m Bromides. The isosbestic points made by the intersection of the spectrum of salt with the spectrum of pure water are highlighted by the arrows. More spectra are shown in the supplemental material.

A family of straight lines is obtained by plotting the $\omega_{ip}$ versus the difference between the electric fields of anions and cations $\left(\frac{q}{r_{Anion}^2} - \frac{q}{r_{Cation}^2}\right)$, see Fig. 2. This fact confirms that electrostatic forces are the leading forces not only in the solution formation but also in the re-structuring of water due to the solutes. The higher the difference the greater is $\omega_{i.p.}$ and also the disordering effect of the solute on water because high order-low density fraction is depressed. All the lines, with the exception of the Fluorides one, form a bundle crossing the y-axis at about 3350 cm$^{-1}$. We can thus obtain an equation describing $\omega_{ip}$:

$$\omega_{i.p.} = \frac{A \cdot m}{(e_{Oxy} - e_{Anion})} \left(\frac{q}{r_{Anion}^2} - \frac{q}{r_{Cation}^2}\right) - B \cdot \frac{q}{r_{Anion}^4} + 3391 \qquad (1)$$

where $A = 2,3 \pm 0,1 \ [cm \cdot C^{-1}]$ and $B = 6,7 \pm 0,5 (\cdot 10^5) \ [cm^{-1}]$ are two phenomenological constants obtained by the fit, $m$ is the anion molality and $q$ the charge. The first term represents the difference between the electric fields of the ion and its counter-ion. In the limit where the two fields are equal, the isosbestic point only depends on the anion radius. The first term is multiplied by a factor taking into account the (inverse) difference between the electronegativity of oxygen and anion. Electronegativity is the force of attraction of an atom for bonding electrons, fluorine is the only element which attracts electrons stronger than oxygen and we may guess that the disordering effect of solutes is increased by the effect of fluorine on the lone pair electrons of the OH-bond. This justify the negative slope of fluoride its $\omega_{ip}$ much lower than the other halides. The second term strongly depends on the anion size and it is upper bounded by the value 3391 cm$^{-1}$ obtained by fitting the data. Summarizing: the difference between the electrostatic forces sets the total amount of water molecules stretched or shortened by the effect of the solute.

In table I are shown the value used in the calculations and the comparison between the isosbestic points estimated with the (1) and the measured values. Equation (1) also matches the results for bi-valent ions, in that case the Anion molality has to be taken into account setting the factor $m = 2$. The match between calculated and measured data is very reasonable.

TABLE I

|  | $r_{Anion}$ (nm) | $r_{Cation}$ (nm) | $\frac{1}{r^2_{Anion}} - \frac{1}{r^2_{Cation}}$ | $e_{Oxygen} - e_{Cation}$ | m | i.p. estimated (cm$^{-1}$) | i.p. measured (±4 cm$^{-1}$) |
|---|---|---|---|---|---|---|---|
| LiF | 133 | 76 |  | -0.5 | 1 | 3230±2 | 3224 |
| NaF | '' | 102 |  | '' | 1 | 3193±1 | 3189 |
| KF | '' | 138 |  | '' | 1 | 3169±0 | 3167 |
| LiCl | 181 | 76 | -14.26 | 0.5 | 1 | 3260±3 | 3262 |
| NaCl | '' | 102 | -6.56 | '' | 1 | 3297±1 | 3310 |
| KCl | '' | 138 | -2.20 | '' | 1 | 3318±0 | 3326 |
| RbCl | '' | 152 | -1.28 | '' | 1 | 3322±0 | 3330 |
| CsCl | '' | 167 | -0.53 | '' | 1 | 3325±0 | 3348 |
| CaCl$_2$ | '' | 100 | -16.95 | '' | 2 | 3167±7 | 3159 |
| SrCl$_2$ | '' | 126 | -9.55 | '' | 2 | 3237±4 | 3248 |
| BaCl$_2$ | '' | 142 | -6.87 | '' | 2 | 3263±3 | 3265 |
| LiBr | 196 | 76 | -14.71 | 0.7 | 1 | 3294±2 | 3300 |
| NaBr | '' | 102 | -7.01 | '' | 1 | 3320±1 | 3319 |
| KBr | '' | 138 | -1.79 | '' | 1 | 3338±0 | 3333 |
| LiI | 220 | 76 | -15.24 | 1 | 1 | 3324±2 | 3318 |
| NaI | '' | 102 | -7.54 | '' | 1 | 3342±1 | 3333 |
| KI | '' | 138 | -2.32 | '' | 1 | 3354±0 | 3347 |
| RbI | '' | 152 | -1.79 | '' | 1 | 3355±0 | 3350 |

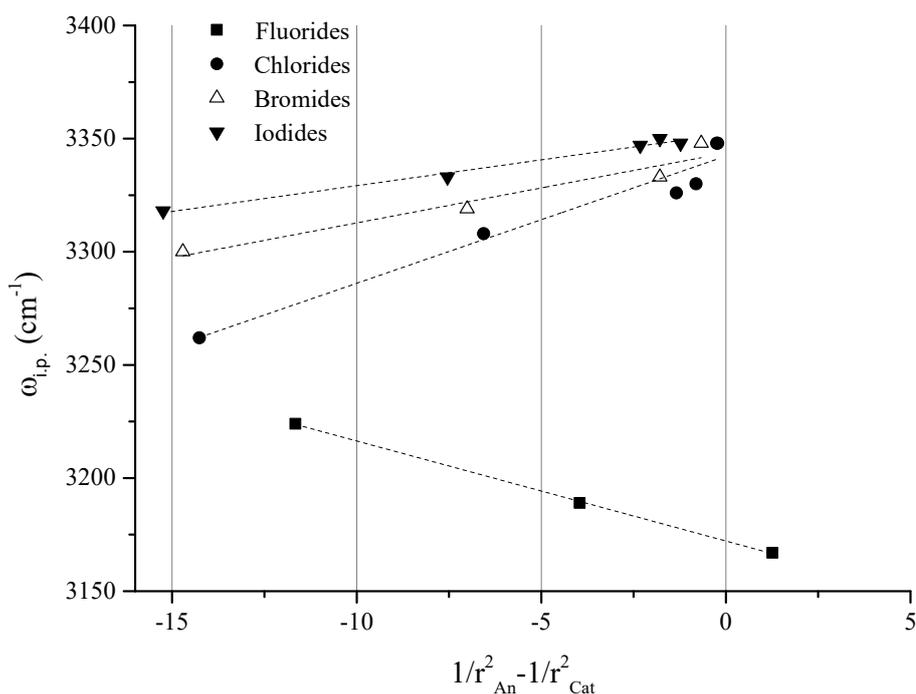

Fig.2 Isosbestic points versus the difference between the electrostatic forces of ion and its counter-ion for monovalent alkali halides.

Equation (1) also correctly describes the state of a mixture of two equal concentrations of solutions having an anion common. For instance mixing two solutions 1m NaCl and CsCl a new isosbestic point is obtained whose abscissa is the average of that of the two components. This feature reveals the addictive property of the isosbestic points as can be expected by its electrostatic nature. In particular, in case of a mixture of two salts of the same halide, the appearance of the isosbestic point is due to the balance of the electrostatic force between the anion and a cation having an apparent ionic radius equal to the average of the ionic radii of the two cations.

We also used equation (1) to evaluated the ionic radius of the monovalent ion $NH_4^+$ of ammonium bromide. We have measured $\omega_{i.p.}$ = 3348 cm$^{-1}$ in line with other bromides and corresponding to an effective radius r = 134 pm to be compared with the value stated in the literature r = 137 pm [13].

DETERMINATION OF DENSITY CHANGES IN SOLUTIONS

The density of electrolytic solutions has found to increase linearly with the concentration because of the replacement of two water molecules with heavier couples of ions [14]. The slopes of the linear trend depend both on the ion and cation masses (details are available in the Supplemental Material). To understand the role of the water molecules re-arrangement in solutions we have studied the change of the density with the concentration of the alkali halides and we have found that, being $c$ the molarity, the rate $\frac{\partial \rho}{\partial c}$ is related to the isosbestic point of the vibrational spectra of water as shown in Table II and in Fig.3. This fact is not unexpected because it is known that the presence of ions changes the specific volume of the water fraction of the solution [15]. The isosbestic point actually describes how water molecules modify their OH bond length in order to face the complex electrostatic force balance due to the solute.

Data are easily fitted by with the equation:

$$\frac{\partial \rho}{\partial c} = C \cdot \left(\omega_{i.p.} - \omega_0\right)^2 + D \qquad (2)$$

Where $C = (8 \pm 2) \cdot 10^{-6}$, $D = 0.01 \pm 0.01$ and $\omega_0 = 3236 \pm 8$ cm$^{-1}$ are parameters of the fit. We can thus observe that the density of an electrolytic solution also depends by the order/disorder degree of water induced by the electrolyte on the long-range structure of the liquid.

## Table II

Density data are from A.V. Wolf, <u>Aqueous Solutions and Body Fluids</u>, Hoeber, 1966

| Sali | dρ/dc | $\omega_{i.p.}$ exp |
|---|---|---|
| LiCl | 0.0239 | 3262 |
| LiBr | 0.0623 | 3300 |
| LiI | 0.0972 | 3318 |
| NaF | 0.0406 | 3189 |
| NaCl | 0.0406 | 3310 |
| NaBr | 0.0788 | 3319 |
| NaI | 0.1123 | 3333 |
| KF | 0.0494 | 3167 |
| KCl | 0.0453 | 3326 |
| KBr | 0.0828 | 3333 |
| KI | 0.1157 | 3347 |
| RbF | 0.0866 | |
| RbCl | 0.0908 | 3330 |
| RbBr | 0.1224 | |
| RbI | 0.1541 | 3350 |
| CsF | 0.1250 | |
| CsCl | 0.1245 | 3348 |
| CsBr | 0.1594 | |
| CsI | 0.1893 | |

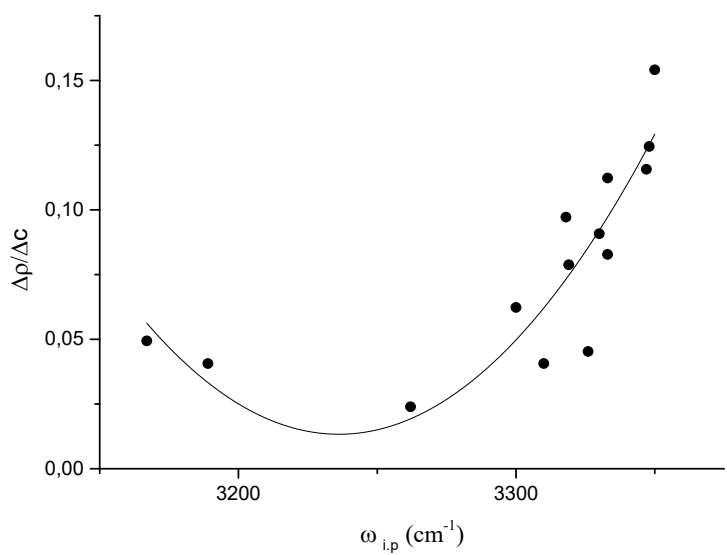

Fig. 3 Rate change of the density of alkali halides versus the measured isosbestic point. On y-axis we have represented the slopes α of the straight lines $\rho = \alpha c + \rho_0$ where c is the molarity and $\rho_0$ (=1) is the pure water density. The black line is the fit with eq. (2).

# AQUEOUS HYDROGENS AND HYDRATED HYDRONIUM

Proton transfer in aqueous media has been the subject of many experimental and theoretical studies both in chemistry and in biology because of its key role in many functions. The hydrated proton shows a very high mobility compared to the other species. The commonly accepted model of proton transfer is that the excess positive charge migrates from one water molecule to the next one in a structural process termed as Grotthus mechanism [16]. $H^+$ is unlike the other ions in water: its properties differ dramatically. It cannot exist in solutions because of its very high charge density and forms hydronium ion $H_3O^+$ that has a flattened trigonal pyramidal geometry and an effective radius ≈ 100 pm [17] even lower than the water molecular radius $r_w$ = 138 pm. Actually the existence of $H_3O^+$ has only been observed with X-rays performed on solids acid hydrates like nitric acid or perchloric acid, however, it has been assumed that the ions existing in one phase survives the phase transition without structural changes [18]. In order to make the assignment of the IR spectrum of aqueous acids solution it has been also assumed the existence of larger protonated water clusters. The hydronium ion binds strongly to another water molecule and may form double hydrogen bounded Zundel $H_5O_2^+$ ions, triple hydrogen bounded Eigen $H_9O_4^+$ ions or even larger clusters [19, 20]. Even halogenic acids show an isosbestic point with the increase of the concentration, however, if we try to apply the equation (1) with $r_{Cation}$= 100 pm we obtain points far out the corresponding halides straight line, see fig.2. As it can be seen in Fig. 4, the crossing point is at about 3142 $cm^{-1}$ out of the range 3200 - 3400 $cm^{-1}$ observed for aqueous solutions of salts. Actually, in case of halogenic acids, the isosbestic point is non-exact: i.e. the spectra crosses in about 20 $cm^{-1}$ wide range. This feature implies a weak concentration dependence of $\omega_{i.p.}$ [21]. We suggest that, in contrast to the role of the isosbestic point in salts solutions where it marks the separation of the region where water-water interaction prevails on the water-ion one, in the case of acids it separates the stretching of hydronium by the region of hydrated hydronium, i.e. Eigen

$H_9O_4^+$ and Zundel $H_5O_2^+$ ions [22]. The growth of a continuum-like absorption band when protons are added to water below $\approx 3000 cm^{-1}$ has been assigned to Zundel stretch (3100-299) and to Eigen stretch (2900-2600 $cm^{-1}$) [23]. Since the isosbestic point defines the existence of a rule of sum , i.e. one population grow at the expense of the other, we guess that water molecules seized to hydrate protons are subtracted to the more correlated water fraction and not randomly taken among all the bulk molecules.

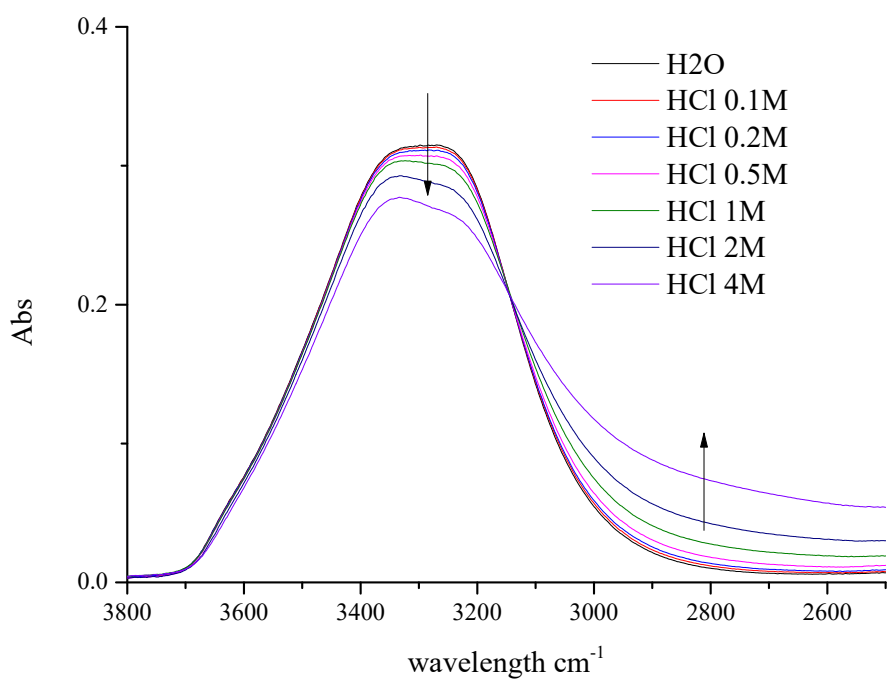

Fig.4 OH stretching band of aqueous solutions of hydrochloric acid. Isosbestic points of halogenic acid are confined in the region 3131–3151 $cm^{-1}$ .

In order to apply the equation (1) the radius $r_{Cation}$ should be known. However, a measure of the effective radius of ion in a compound is quite complex because it has to keep into account that nuclei are surrounded by electron clouds. Moreover, the proton moves quickly by jumping one molecular length at a time. This makes very hard to define its molecular radius because it spends part of its time tightly bound to a water molecule (forming $H_3O^+$) and part between two different molecules before the next bond. We can thus define a Dynamic Effective Radius (DER) as the average between these two configurations, such a value will be affected by the complex electrostatic configuration of the charges in the solution. In case of HCl, the measured isosbestic point is equal to 3148 $cm^{-1}$ and it fits the straight line of the chlorides (see Fig.2) assuming a DER equal to 52 pm. Likewise the isosbestic point of HF fits the fluorides line at DER = 55 pm. HBr and HI gives, respectively, DER = 35 pm and DER = 31 pm. We guess that these findings provide a measure of the binding force: we can define the DER as a weighted average of the radius of $H_3O^+$ and the radius of a free proton: $DER = \alpha \cdot r_{H_3O^+} + (1-\alpha) \cdot r_{proton}$ where $\alpha$ represent the fraction of time spent as $H_3O^+$ bounded to a water molecule. The attraction between the proton and the water molecule also depends on the competition between the oxygen and the anion in solution; ions like $Br^-$ and $I^-$ with a low charge density allow the proton to spend more time unbounded before to be captured by another water molecule, thus showing a lower values of $\alpha$.

In table III are summarized isosbestic points and the dynamic effective radii for binary acids. It is worthwhile to observe that there is a linear relation between the DER and the force of the acid $pK_a$, see figure 5, thus confirming the relationship between the time spent as $H_3O^+$ bounded to a water molecule dynamic effective radius and the hydrated proton. Also it is observed in Table III that the low $K_a$ of HF set the behaviour of the fluoridic acid out of the line of the other halides confirming the role of its high electronegativity in seizing the proton; it is also possible that un-dissociated ionic couples $H^+F^-$ and $H_2F_2$ dimers may affect this finding.

TABLE III

|     | Isosbestic point | Dynamic Effective Radius (DER) | pKa |
|-----|------------------|-------------------------------|------|
| HF  | 3296 ± 8         | 55                            | 3.2  |
| HCl | 3148 ± 3         | 52                            | -7   |
| HBr | 3136 ± 4         | 35                            | -9   |
| HI  | 3132 ± 4         | 31                            | -10  |

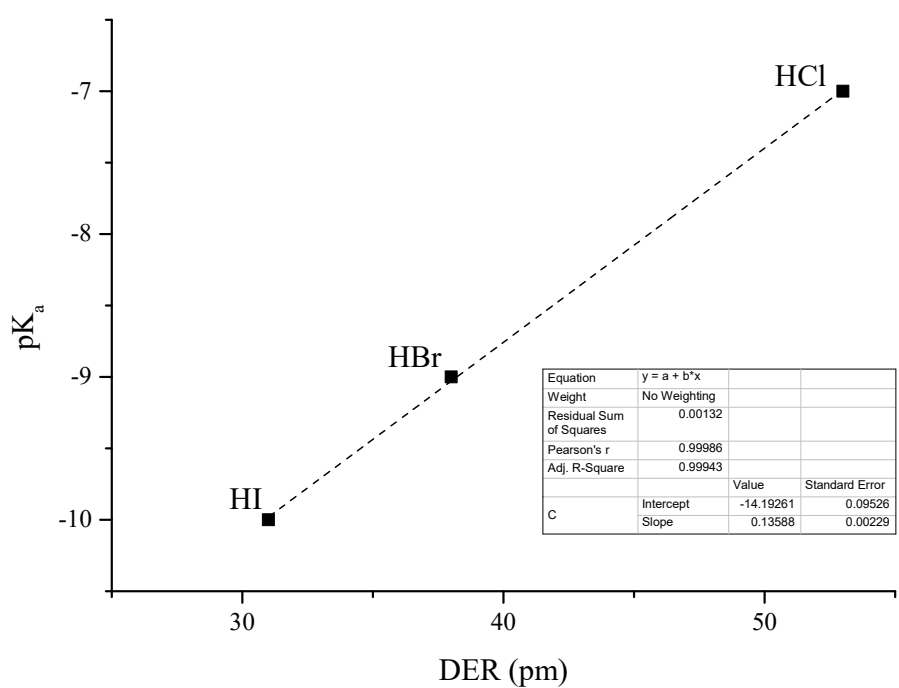

Fig. 5 Acid dissociation constant versus the Dynamic Effective Radius of hydrated protons in aqueous solutions of halogenic acids. The dashed line is the linear fit whose parameters are shown in the insert.

CONCLUSIONS

We have shown as the study of the isosbestic points of the electrolytic solutions actually make available a large amount of information. Herein we have faced three main issues: a) the balance of electrostatic forces acting into the solution; b) the effect of solutes on the average volume occupied by a water molecule; c) the time spent by a proton tightly bound to a water molecule, forming $H_3O^+$, before jumping to the next bond.

The isosbestic point appears as the equilibrium point between two fraction of water molecules whose structure is affected by the solute in a different way. It implies that one population is depleted whilst the other is increased by the addition of the solute and define the crossing point between two fractions: one where water-water interactions prevail on ion-water interactions and the other one where the ion-water interactions are stronger. Vice-versa, just looking at the electrostatic properties of the solutes we have a clue on the long-range structure of water. These results reinforce the image of water as a two fluids system. One fluid, characterised by a strong intermolecular interaction, is destroyed by the presence of a solute in favour of higher-energy, less correlated molecules strongly interacting with the ions in solution. A closely review of the FTIR spectra shows that the interaction with the solutes changes the O-H bond length in relation to the concentration, i.e. with the electrostatic forces due to the complex local electrostatic fields. Accordingly, this fact contributes to determine the density of the solutions.

We have also discussed the case of acidic halides showing that the strong electrostatic forces due to hydrated protons still prevail on the water-water attraction. The isosbestic point observed on acids spectra, in fact, has a different nature: it is the edge between strongly correlated $H_2O$ molecules and protonated water clusters. The existence of an isosbestic point in this wavelength range supports that the water molecules used to form the latter are subtracted to large cluster of water molecules. On the opposite, should they be randomly distributed in the bulk a FTIR spectra shift, instead an isosbestic point, should be seen due to the increase of the hydrated protons in the solution. We have also shown that the concept of the static size of hydrogen and hydronium ions has to be reconsidered in the light of the very fast motion of protons between two water molecules. These findings can also shed some light on the question why the size of hydronium ions appear smaller than that of the size of a water molecule.

In conclusion, we want to suggest that the method shown in this paper may also be applied in many other examples of binary halides compounds such as, for example, the ionic liquids, to evaluate the size of unknown cations.

**Acknowledgments**

We thank A. Pozio, E. Simonetti and G. Appetecchi for reading the manuscript and for many useful discussions and suggestions.

This work was supported by the Italian National Agency for New technologies, Energy and Sustainable Economic development.

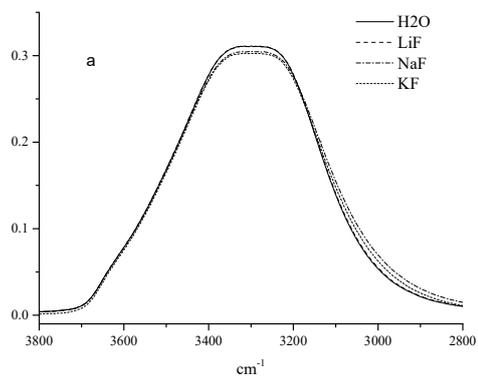
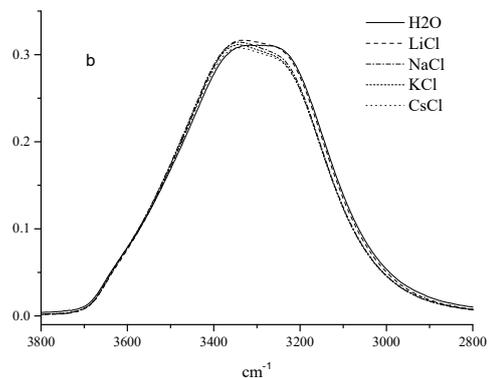
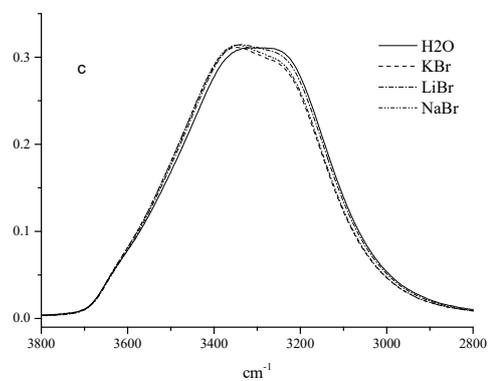
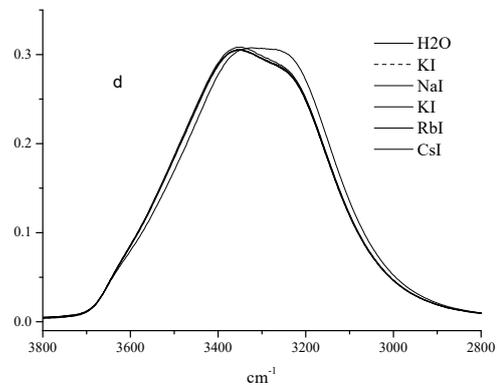
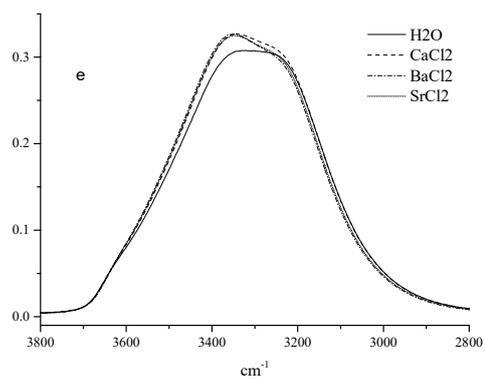

FTIR spectra of halides: a) Fluorides; b) Chlorides; c) Bromides; d) Iodides; e) bi-valent Chlorides.

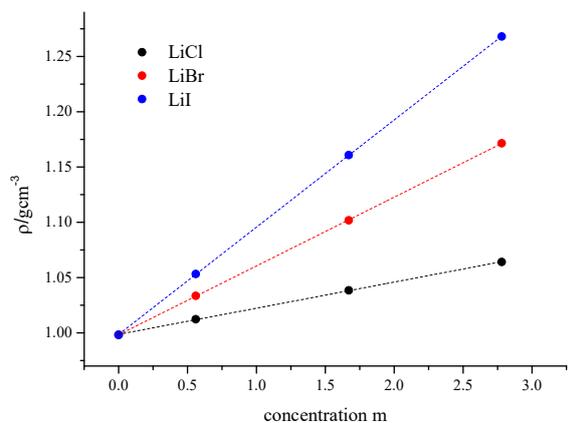
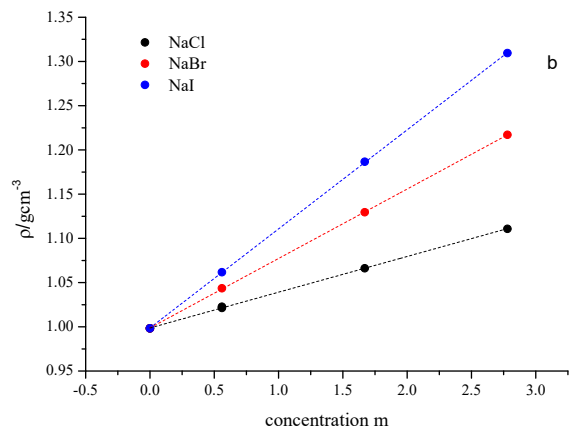
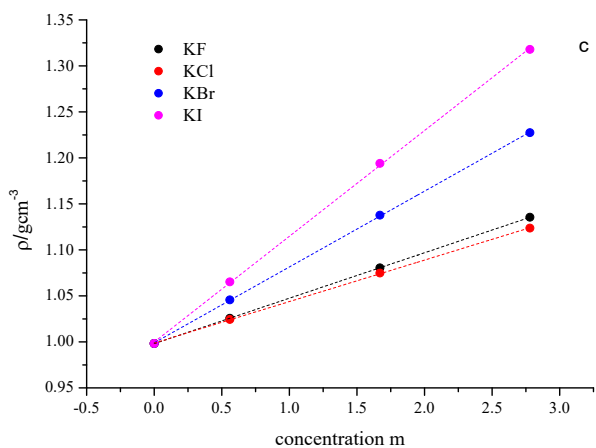
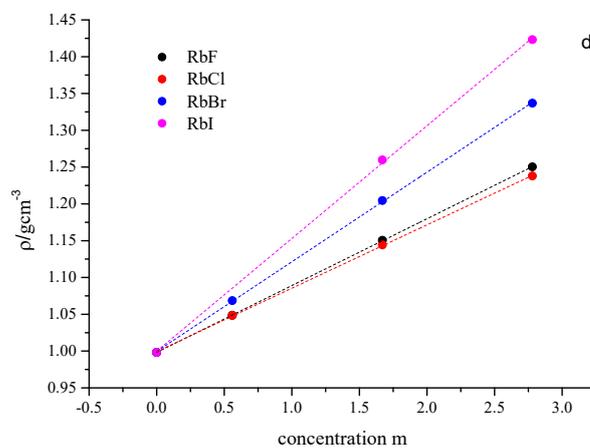
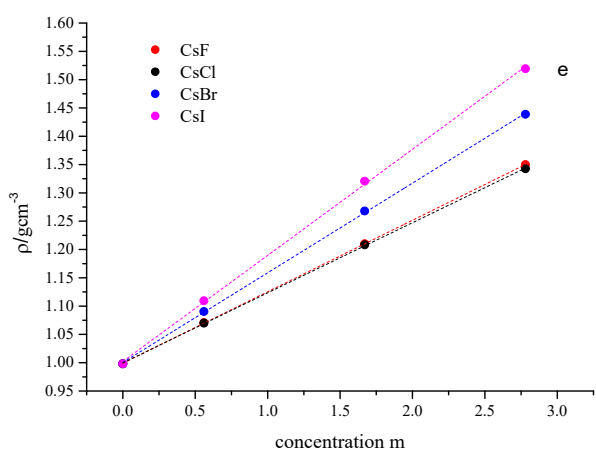

Density changes versus molar concentration for alkali halide: a) Lithium salts; b) Sodium salts; c) Potassium salts; d) Rubidium salts; e) Caesium salts. Data are from
https://chemistry.mdma.ch/hiveboard/rhodium/pdf/chemical-data/prop_aq.pdf